\documentstyle[12pt,psfig]{article}
\catcode`\@=11

\newcommand{\nsect}{\setcounter{equation}{0}
\def\theequation{\thesection\arabic{equation}}\section}

\def\NPB#1#2#3{{\it Nucl.~Phys.} {\bf{B#1}} (19#2) #3}
\def\PLB#1#2#3{{\it Phys.~Lett.} {\bf{B#1}} (19#2) #3}
\def\PRD#1#2#3{{\it Phys.~Rev.} {\bf{D#1}} (19#2) #3}

\def\AP#1#2#3{{\it Ann.~Phys.} {\bf#1} (19#2) #3}

\def\@normalsize{\@setsize\normalsize{15pt}\xiipt\@xiipt
\abovedisplayskip 14pt plus3pt minus3pt%
\belowdisplayskip \abovedisplayskip
\abovedisplayshortskip  \z@ plus3pt%
\belowdisplayshortskip  7pt plus3.5pt minus0pt}
\def\small{\@setsize\small{13.6pt}\xipt\@xipt
\abovedisplayskip 13pt plus3pt minus3pt%
\belowdisplayskip \abovedisplayskip
\abovedisplayshortskip  \z@ plus3pt%
\belowdisplayshortskip  7pt plus3.5pt minus0pt
\def\@listi{\parsep 4.5pt plus 2pt minus 1pt
            \itemsep \parsep
            \topsep 9pt plus 3pt minus 3pt}}

\def\underline#1{\relax\ifmmode\@@underline#1\else
        $\@@underline{\hbox{#1}}$\relax\fi}
\@twosidetrue
\relax

\catcode`@=12

\catcode`\@=11

\def\section{\@startsection{section}{1}{\z@}{3.5ex plus 1ex minus
   .2ex}{2.3ex plus .2ex}{\large\bf}}
\def\thesection{\arabic{section}.}

\def\ps@headings{\def\@oddfoot{}\def\@evenfoot{}
\def\@oddhead{\hbox{}\hfill
        \makebox[.5\textwidth]{\raggedright\ignorespaces --\thepage{}--
        \hfill }}
\def\@evenhead{\@oddhead}
\def\subsectionmark##1{\markboth{##1}{}}
}

\ps@headings

\catcode`\@=12

\relax

\def\figcap{\section*{Figure Captions\markboth
        {FIGURECAPTIONS}{FIGURECAPTIONS}}\list
        {Fig. \arabic{enumi}:\hfill}{\settowidth\labelwidth{Fig. 999:}
        \leftmargin\labelwidth
        \advance\leftmargin\labelsep\usecounter{enumi}}}
 \relax
\def\tablecap{\section*{Table Captions\markboth
        {TABLECAPTIONS}{TABLECAPTIONS}}\list
        {Table \arabic{enumi}:\hfill}{\settowidth\labelwidth{Table 999:}
        \leftmargin\labelwidth
        \advance\leftmargin\labelsep\usecounter{enumi}}}
 \relax
\def\reflist{\section*{References\markboth
        {REFLIST}{REFLIST}}\list
        {[\arabic{enumi}]\hfill}{\settowidth\labelwidth{[999]}
        \leftmargin\labelwidth
        \advance\leftmargin\labelsep\usecounter{enumi}}}
 \relax

\catcode`\@=11

\def\marginnote#1{}
\newcount\hour
\newcount\minute
\newtoks\amorpm
\hour=\time\divide\hour by60
\minute=\time{\multiply\hour by60 \global\advance\minute by-
\hour}
\edef\standardtime{{\ifnum\hour<12 \global\amorpm={am}%
    \else\global\amorpm={pm}\advance\hour by-12 \fi
    \ifnum\hour=0 \hour=12 \fi
    \number\hour:\ifnum\minute<100\fi\number\minute\the\amorpm}}
\edef\militarytime{\number\hour:\ifnum\minute<100\fi\number\minute}
\def\draftlabel#1{{\@bsphack\if@filesw {\let\thepage\relax
  \xdef\@gtempa{\write\@auxout{\string
    \newlabel{#1}{{\@currentlabel}{\thepage}}}}}\@gtempa
    \if@nobreak \ifvmode\nobreak\fi\fi\fi\@esphack}
     \gdef\@eqnlabel{#1}}
\def\@eqnlabel{}
\def\@vacuum{}
\def\draftmarginnote#1{\marginpar{\raggedright\scriptsize\tt#1}}
\def\draft{\oddsidemargin -.5truein
        \def\@oddfoot{\sl preliminary draft \hfil
        \rm\thepage\hfil\sl\today\quad\militarytime}
        \let\@evenfoot\@oddfoot \overfullrule 3pt
        \let\label=\draftlabel
        \let\marginnote=\draftmarginnote
   
\def\@eqnnum{(\theequation)\rlap{\kern\marginparsep\tt\@eqnlabel}%
\global\let\@eqnlabel\@vacuum}  }
\def\preprint{\twocolumn\sloppy\flushbottom\parindent 1em
        \leftmargini 2em\leftmarginv .5em\leftmarginvi .5em
        \oddsidemargin -.5in    \evensidemargin -.5in
        \columnsep 15mm \footheight 0pt
        \textwidth 250mmin      \topmargin  -.4in
        \headheight 12pt \topskip .4in
        \textheight 175mm
        \footskip 0pt
        
\def\@oddhead{\thepage\hfil\addtocounter{page}{1}\thepage}
        \let\@evenhead\@oddhead \def\@oddfoot{} \def\@evenfoot{} 
}
\def\titlepage{\@restonecolfalse\if@twocolumn\@restonecoltrue\onecolumn
     \else \newpage \fi \thispagestyle{empty}\c@page\z@
        \def\thefootnote{\fnsymbol{footnote}} }
\def\endtitlepage{\if@restonecol\twocolumn \else  \fi
        \def\thefootnote{\arabic{footnote}}
        \setcounter{footnote}{0}}  
\catcode`@=12
\relax

\def\ps@headings{\def\@oddfoot{}\def\@evenfoot{}
\def\@oddhead{\hbox{}\hfill
        \makebox[.5\textwidth]{\raggedright\ignorespaces --\thepage{}--
        \hfill }}
\def\@evenhead{\@oddhead}
\def\subsectionmark##1{\markboth{##1}{}}
}

\ps@headings

\relax

\def\firstpage#1#2#3#4#5#6{
\begin{document}
\begin{titlepage}
\nopagebreak
\title{\begin{flushright}
        \vspace*{-1.8in}
        {\normalsize CERN-TH/97-90}\\[-9mm]
        {\normalsize CPTH-S504.0497}\\[-9mm]
        {\normalsize IEM-FT-154/97}\\[-9mm]
        {\normalsize hep-th/9705037}\\[4mm]
\end{flushright}
\vfill
{#3}}
\author{\large #4 \\[1.0cm] #5}
\maketitle
\vskip -7mm     
\nopagebreak 
\begin{abstract}
{\noindent #6}
\end{abstract}
\vfill
\begin{flushleft}
\rule{16.1cm}{0.2mm}\\[-3mm]
$^{\star}${\small Research supported in part by\vspace{-4mm}
IN2P3-CICYT under contract Pth 96-3, in part by 
CICYT contract AEN95-0195, and
in part by the EEC under the TMR contract \vspace{-4mm}
ERBFMRX-CT96-0090 and under CHRX-CT93-034.}\\[-3mm] 
$^{\dagger}${\small Laboratoire Propre du CNRS UPR A.0014.}\\
CERN-TH/97-90\\
April 1997
\end{flushleft}
\thispagestyle{empty}
\end{titlepage}}

\def\simlt{\stackrel{<}{{}_\sim}}
\def\simgt{\stackrel{>}{{}_\sim}}
\newcommand{\dal}{\raisebox{0.085cm}
{\fbox{\rule{0cm}{0.07cm}\,}}}
\newcommand{\dt}{\partial_{\langle T\rangle}}
\newcommand{\dtbar}{\partial_{\langle\bar{T}\rangle}}
\newcommand{\al}{\alpha^{\prime}}
\newcommand{\mst}{M_{\scriptscriptstyle \!S}}
\newcommand{\mpl}{M_{\scriptscriptstyle \!P}}
\newcommand{\dv}{\int{\rm d}^4x\sqrt{g}}
\newcommand{\lv}{\left\langle}
\newcommand{\rv}{\right\rangle}
\newcommand{\ph}{\varphi}
\newcommand{\abar}{\bar{a}}
\newcommand{\sbar}{\,\bar{\! S}}
\newcommand{\xbar}{\,\bar{\! X}}
\newcommand{\fbar}{\,\bar{\! F}}
\newcommand{\zbar}{\bar{z}}
\newcommand{\dbar}{\,\bar{\!\partial}}
\newcommand{\tbar}{\bar{T}}
\newcommand{\taubar}{\bar{\tau}}
\newcommand{\ubar}{\bar{U}}
\newcommand{\ybar}{\bar{Y}}
\newcommand{\phb}{\bar{\varphi}}
\newcommand{\cm}{Commun.\ Math.\ Phys.~}
\newcommand{\prl}{Phys.\ Rev.\ Lett.~}
\newcommand{\pr}{Phys.\ Rev.\ D~}
\newcommand{\pl}{Phys.\ Lett.\ B~}
\newcommand{\ibar}{\bar{\imath}}
\newcommand{\jbar}{\bar{\jmath}}
\newcommand{\np}{Nucl.\ Phys.\ B~}
\newcommand{\F}{{\cal F}}
\renewcommand{\L}{{\cal L}}
\newcommand{\A}{{\cal A}}
\newcommand{\e}{{\rm e}}
\newcommand{\be}{\begin{equation}}
\newcommand{\en}{\end{equation}}
\newcommand{\gsi}{\,\raisebox{-0.13cm}{$\stackrel{\textstyle
>}{\textstyle\sim}$}\,}
\newcommand{\lsi}{\,\raisebox{-0.13cm}{$\stackrel{\textstyle
<}{\textstyle\sim}$}\,}
\date{}
\firstpage{3118}{IC/95/34}
{\large\bf Supersymmetry breaking in M-theory and gaugino condensation} 
{I. Antoniadis$^{\,a,b}$ and M. Quir{\'o}s$^{\,c}$}
{\normalsize\sl
$^a$TH-Division, CERN, CH-1211 Geneva 23, Switzerland\\[-3mm]
\normalsize\sl$^b$ Centre de Physique Th{\'e}orique, 
Ecole Polytechnique,$^\dagger$ {}F-91128 Palaiseau, France\\[-3mm]
\normalsize\sl
$^c$Instituto de Estructura de la Materia, CSIC, Serrano 123, 28006 Madrid,
Spain}
{ We argue that supersymmetry breaking by gaugino condensation in the 
strongly coupled heterotic string can be described by an analogue of 
Scherk--Schwarz compactification on the eleventh dimension in M-theory.
The M-theory scale is identified with the gauge coupling unification mass,
whereas the radius of the eleventh dimension $\rho$ is at an intermediate scale
$\rho^{-1}\sim 10^{12}$ GeV. At the lowest order, supersymmetry is broken only
in the gravitational and moduli sector at a scale $m_{3/2}\sim\rho^{-1}$,
while it is mediated by gravitational interactions to the observable world.
Computation of the mass splittings yields in general a
hierarchy of soft masses at the TeV scale $(\sim\rho^{-2}/M_p)$
with matter scalars much heavier than gauginos.}

\nsect{Introduction}

It is now widely believed that the strongly coupled 
$E_8\times E_8$ heterotic string theory is described by M-theory whose 
low energy limit is the eleven-dimensional supergravity~\cite{hw}. 
In particular, four-dimensional (4D) $N=1$
supersymmetric vacua of the heterotic string compactified on Calabi--Yau (CY)
manifolds are described, in the strong 10D coupling regime, by M-theory
compactifications on CY$\times S^1/Z_2$, the radius $\rho$ of the 
semicircle $S^1/Z_2$ being large~\cite{w}. 
In view of this duality, it is of the highest interest to discuss outstanding 
problems of perturbative string theory, 
such as supersymmetry breaking~\cite{h}.

A first consequence of heterotic -- M-theory duality is on gauge coupling
unification~\cite{w,aq}. 
By identifying the M-theory scale with the unification mass $\sim 10^{16}$ GeV, 
it was noticed that the threshold of the eleventh dimension $\rho^{-1}$ is
at an intermediate scale $\sim 10^{12}$ GeV, while the inverse size of
the CY manifold (in the isotropic case) should also be of the order of the 
unification scale~\cite{aq}. 
{}Furthermore, it was proposed to relate this intermediate
scale with the scale of supersymmetry breaking. More precisely, when
supersymmetry is broken on the M-theory side by a coordinate-dependent 
compactification analogue to the Scherk--Schwarz 
mechanism~\cite{ss} using the
eleventh dimension, one obtains that to lowest order only the
gravitational sector is affected with a gravitino mass 
$m_{3/2}\sim\rho^{-1}$. The observable world, living at the end-points of the
semicircle, remains unaffected and feels supersymmetry breaking only by
gravitational interactions, which yield a hierarchy of values  
for the soft breaking terms in the TeV range,
$m_{\rm susy}\sim\rho^{-2}/M_p$. This therefore provides 
an appealing and economical scenario, 
which relates unification and supersymmetry breaking scales. Moreover, it was
suggested that this situation could describe ordinary gaugino 
condensation in the strongly coupled heterotic string~\cite{aq}.

In this work, we make the above analogy with the heterotic string 
precise and perform a detailed analysis of supersymmetry breaking in M-theory. 
In section 2, we
describe the compactification of M-theory on CY$\times S^1/Z_2$ and discuss
duality relations and the unification condition. In section 3, we review the
main features of supersymmetry breaking by gaugino condensation in the heterotic
string. Comparison with the duality results motivates the relation 
$m_{3/2}\sim 10^{12}\,{\rm GeV}\sim\rho^{-1}$, which can be obtained by a
coordinate-dependent compactification on the eleventh dimension. This mechanism
is made explicit in section 4. In section 5, we compute the effects of
supersymmetry breaking in the observable sector. We find that all scalar masses
arise in general by gravitational interactions at the one-loop level as
$\sim m_{3/2}^2/M_p$, while the dominant contributions to gaugino masses 
arise at the next level by gauge interactions as 
$\sim(\alpha/2\pi)\, m_{3/2}^2/M_p$.
{}Finally, section 6 contains our conclusions and outlook.

\nsect{Compactification of M-theory on CY$\times$S$^1$/Z$_2$}

Here, we review the main properties of M-theory compactification in four
dimensions on a seven-dimensional internal space, which is the product of a 
Calabi--Yau manifold with the semicircle $S^1/Z_2$. Proceeding in 
two steps, we will first consider the five-dimensional theory on a Calabi--Yau
threefold with Hodge numbers $h_{(1,1)}$ and $h_{(2,1)}$ leading to $N=1$ 5D
space-time supersymmetry~\cite{ccaf}. 
In addition to the gravitational multiplet, 
the massless spectrum consists of $n_V=h_{(1,1)}-1$ vector multiplets and
$n_H=h_{(2,1)}+1$ hypermultiplets. The gauge group is abelian, $U(1)^{n_V+1}$,
where the additional factor counts the graviphoton.
Starting with the eleven-dimensional bosonic
fields, which are the metric $g$ and the 3-form gauge potential $A$, and
splitting the Lorentz indices as $(\hat\mu,i,\bar\jmath)$ with 
$\hat\mu=1,\dots,5$ and $i,\bar\jmath=1,2,3$, 
the $h_{(1,1)}$ gauge fields are given by 
$A_{\hat\mu i\bar\jmath}$ while the $h_{(1,1)}-1$ vector moduli correspond to 
$g_{i\bar\jmath}$ with unit determinant. Moreover, the hypermultiplet moduli
are given by the $h_{(2,1)}$ complex scalar pairs $(g_{ij},A_{ij\bar{k}})$,
along with the universal scalars $(\det g_{i\bar\jmath},
A_{\hat\mu\hat\nu\hat\lambda},A_{ijk}=a\epsilon_{ijk})$.

The second step consists in the compactification of the previous 5D theory
down to four dimensions on $S^1/Z_2$, where the $Z_2$ acts as an inversion on
the fifth coordinate $x_5\to -x_5$ and changes the sign of the 11D 3-form
potential $A\to -A$~\cite{hw}. 
As a result, one obtains $N=1$ supersymmetry in four
dimensions together with $h_{(1,1)}+h_{(2,1)}+1$ massless chiral multiplets. The
corresponding scalar moduli are the $h_{(1,1)}$ real pairs $(g_{i\bar\jmath},
A_{5i\bar\jmath})$, the $h_{(2,1)}$ complex scalars $g_{ij}$ and the universal
real pair $(g_{55},A_{5\mu\nu})$. 

On top of the massless states, there is the
usual tower of their Kaluza--Klein excitations with masses
\be
M^2={n^2\over\rho^2}\quad ;\quad n=0,\pm 1,\dots
\label{KK}
\en
corresponding to the fifth component of the momentum, $p_5$, which is 
quantized in units of the inverse radius of $S^1$, $1/\rho$. Because of the
$Z_2$ projection, only the symmetric combination of their excitations 
$|n\rangle +|$--$n\rangle$ survive. On the other hand, the $Z_2$-odd states
that were projected away at the massless level, give rise to massive
excitations corresponding to the antisymmetric combination 
$|n\rangle -|$--$n\rangle$. It follows that all states of the 5D theory appear 
at the massive level.

In addition to these untwisted fields, there
are twisted states located at the two end-points of the semicircle, giving
rise to the gauge group and to ordinary matter representations. 
In the case of the
standard embedding, there is an $E_8$ sitting at one of the end-points and an
$E_6$ with $h_{(1,1)}$ ${\bf 27}$ and $h_{(2,1)}$ $\overline{\bf 27}$
matter chiral multiplets sitting at the other. Of course, in any
realistic model, $E_6$ should be further broken down to the Standard Model
gauge group, for instance by turning on (discrete) Wilson lines.

This theory is believed to describe the strong coupling regime of the heterotic
superstring compactified on the same Calabi--Yau manifold. The corresponding 
duality relations that express the eleven-dimensional parameters in terms of
the four-dimensional ones are given by~\cite{w,aq}:
\begin{eqnarray}
M_{11} &=& (2\alpha_G V)^{-1/6}\nonumber\\
\rho^{-1} &=& \left({2\over\alpha_G}\right)^{3/2}M_p^{-2}V^{-1/2}=
{4\over\alpha_G}M_{11}^3 M_p^{-2}\ ,
\label{rel}
\end{eqnarray}
where $\alpha_G$ is the gauge coupling at the unification scale,
$M_p=G_N^{-1/2}$ is the Planck mass, $(2\pi)^6 V$ is the volume of the 
Calabi--Yau space, and $M_{11}$ the mass scale of M-theory. 

The description of the strongly coupled heterotic
string in terms of M-theory is valid in the region where the ten-dimensional
heterotic coupling is large, implying that $\rho$ is much larger than the 
Planck length, or more precisely $\rho\gg 4\alpha_G^{-1/2}M_p^{-1}$.
This condition is held, in particular, when we identify the M-theory scale 
with the unification mass, $M_{11}\sim M_G\simeq 2\times 10^{16}$ GeV, 
inferred from the low-energy data. Equation (\ref{rel}) then 
implies~\cite{aq,llna}:
\be
\rho^{-1}\simeq 10^{12}\ {\rm GeV}\ ,
\label{rho}
\en
while the compactification scale for an isotropic Calabi--Yau manifold is also
of the order of $M_G$, $V^{-1/6}\simeq 1.3\times 10^{16}$ GeV. 

{}From the heterotic point of view, although the 4D coupling is
small $(\alpha_G\sim 1/25)$ there are
strong coupling effects associated to the fact that the 10D
coupling is large, because the compactification scale $V^{-1/6}$ is 
smaller than the heterotic string scale 
$M_H=(\alpha_G/8)^{1/2}M_p\sim 10^{18}$ GeV~\cite{caceres}.
As a result, above the intermediate scale (\ref{rho}) the 4D effective field
theory description breaks down, the ``eleventh" dimension of the M-theory opens
up, and the theory behaves effectively as five-dimensional. However, in this
compactification, only the gravitational and moduli sectors of the theory 
live in the 5D bulk. 
The gauge and matter sectors live on the two 4D boundaries 
associated to the two 
end-points of the semicircle. The ``observable" world lives on one of the
boundaries while the ``hidden" $E_8$ is sitting at the other. They
communicate with each other and with the bulk only through gravitational
interactions.

\nsect{Supersymmetry breaking in the heterotic string by gaugino condensation}

On the heterotic side, one expects that (local) supersymmetry can be broken by
gaugino condensation effects in the hidden $E_8$, at least in the (10D) weakly
coupled regime~\cite{gc1}. Let us describe briefly
the main features of this
mechanism. The physical picture is that the condensate
$\langle\lambda\lambda\rangle$ develops at a scale $\Lambda_c$, where the gauge
coupling of the hidden $E_8$ becomes strong:
\be{\displaystyle
\langle\lambda\lambda\rangle\sim\Lambda_c^3=
\mu^3e^{-{2\pi\over c_8\alpha_8(\mu)}}}\ ,
\label{ll}
\en
with $c_8=30$ being the quadratic Casimir of $E_8$ and $\alpha_8(\mu)$ 
its coupling constant at the scale $\mu$.

This phenomenon can be described by introducing a chiral supermultiplet $U$
whose vacuum expectation value (VEV) reproduces the condensate 
(\ref{ll})~\cite{gc2}. The
effective non-perturbative superpotential is determined by consideration of the
anomalous Ward identities:
\be
W_{\rm np}\propto U\left(
{1\over\alpha_W}+{c_8\over2\pi}\ln{U\over \mu^3}\right)\ ,
\label{w}
\en
where $\alpha_W$ is the Wilsonian effective coupling (at the scale $\mu$), 
which depends holomorphically on the moduli~\cite{sv}. 
It is related to the physical coupling by:
\be
{1\over\alpha_8}={\rm Re}{1\over\alpha_W}+{c_8\over 4\pi}
(-K+2\ln(S+{\bar S}))\ ,
\label{awa8}
\en
where $K$ is the K{\"a}hler potential and $S$ is the heterotic dilaton whose
VEV determines the 4D string coupling constant, Re$S=1/\alpha_G$.

Minimization of the effective potential 
with respect to $U$ implies to leading order in $\Lambda_c/M_p$ the condition 
$\partial_U W_{\rm np}=0$~\cite{bdqq}, which gives
\be{\displaystyle
U=\mu^3e^{-{2\pi\over c_8\alpha_W}-1}\qquad ;\qquad
W_{\rm np}\propto U}\ .
\label{U}
\en
Using this result together with eqs.~(\ref{ll}) and (\ref{awa8}), it is
straightforward to obtain the value of the gravitino mass:
\be
m_{3/2}=|W_{\rm np}|e^{K/2}\propto {1\over\alpha_G}\Lambda_c^3M_p^{-2}\ .
\label{m32}
\en

The effective potential should also be minimized with respect to the dilaton
field $S$. Unfortunately, its runaway behaviour drives the theory to the
supersymmetric limit with vanishing coupling, $S\rightarrow\infty$. A
possible stabilization mechanism was initially proposed by means of a
VEV for the field-strength of the antisymmetric tensor field along the compact
directions, which shifts the superpotential by a constant~\cite{gc1}. However,
this constant was found to be quantized, so that $W_{\rm np}$ becomes of order
one at the minimum~\cite{rw}. Then, eq.~(\ref{m32}) implies that the only way
to obtain a hierarchy for the gravitino mass is by making $e^{K/2}$ small, or
equivalently by having a large compactification volume $V\sim e^{-K}$.
As a result, we obtain the following scaling relations (in $M_p$ units):
\be
m_{3/2}\sim V^{-1/2}\qquad ; \qquad \Lambda_c\sim V^{-1/6}\ .
\label{scaling}
\en

Assuming that eqs.~(\ref{m32}) and (\ref{scaling}) hold in the strong 
coupling regime, a comparison with the duality relations (\ref{rel})
implies the identification of
the condensation scale $\Lambda_c$ with the M-theory scale $M_{11}$
and the inverse radius of the semicircle $\rho^{-1}$ with the gravitino mass:
\be
\Lambda_c\sim M_{11}\qquad ; \qquad m_{3/2}\sim \rho^{-1}\ .
\en

On the other hand, the scale $m_{3/2}$ can be fixed by imposing the 
unification condition $M_{11}\sim M_G$, which is also consistent with the fact
that $\Lambda_c$ is very close to $M_G$, owing to the rapid variation of the
$E_8$ gauge coupling. More precisely, taking $\Lambda_c\sim 10^{16}$ GeV, one
obtains from eq.~(\ref{m32}) $m_{3/2}\sim 10^{12}$ GeV, precisely the
value (\ref{rho}) deduced from unification. This scenario can be
phenomenologically viable if all mass splittings in the observable sector are
zero to ${\cal O}(m_{3/2})$ and are originated by gravitational interactions.
Such a situation was realized in the past~\cite{gcold1,gcold2}, 
in the simplest case where the gaugino condensation
potential is stabilized by the VEV of the antisymmetric tensor field-strength,
because of the particular form of the effective field theory.

The previous arguments suggest that the gaugino condensation mechanism for
supersymmetry breaking in the strongly coupled heterotic string  could be
described on the M-theory side by the field-theoretical  Scherk--Schwarz
mechanism on the eleventh dimension, which generically yields the desired
relation $m_{3/2}\sim\rho^{-1}$. This is our main motivation to study
supersymmetry breaking in M-theory by such coordinate-dependent
compactification, as we describe in the next section.

\nsect{Supersymmetry breaking in M-theory by the Scherk--Schwarz mechanism on
the eleventh dimension}

The Scherk--Schwarz mechanism of supersymmetry breaking makes use of a symmetry
of the theory transforming the gravitino non-trivially. Upon compactification of
the coordinate $x_5$ on a circle of radius $\rho$, one then chooses the
five-dimensional fields to be periodic up to a symmetry transformation,
$\Phi_q(x_5+2\pi\rho)=e^{2i\pi q\omega}\Phi_q(x_5)$, $q$ being the charge
and $\omega$ the transformation parameter. This boundary condition leads to a
shift of the momentum $p_5$ along $x_5$ as $p_5=(n+q\omega)/\rho$, 
which gives rise to mass shifts in the four-dimensional theory by
$q\omega/\rho$. Since the symmetries that transform the gravitino are usually
discrete $R$-symmetries, the parameter $\omega$ is
quantized and the scale of supersymmetry breaking is set by $\rho^{-1}$. 

These symmetries are in fact discrete internal rotations left over after the
compactification. Since in our case the coordinate $x_5$ is compactified on a
semicircle $S^1/Z_2$, consistency of the theory requires that the corresponding 
Noether current must transform with a minus sign under the $Z_2$ orbifold action.
This implies that the symmetry must be a rotation on a plane defined by $x_5$
and one of the internal Calabi--Yau coordinates. {}For a generic Calabi--Yau all
such transformations are broken with the possible exception of a 2$\pi$ rotation.
This rotation acts on the 5D fields as the space-time parity $(-)^{2s}$, which
changes the sign of fermions and leaves bosons invariant. As a result, all
fermions in the bulk receive a uniform shift of their $p_5$ momentum as
$p_5=(n+1/2)/\rho$, while bosons remain untouched. On the other hand, 
twisted states living at the edges of the semicircle have no $p_5$ momenta 
and consequently no shifts. As a result, the general mass formula (\ref{KK}) is
modified to:
\begin{eqnarray}
M^2_{\rm ferm - bulk}&=& {\left(n+{1\over 2}\right)^2\over\rho^2}\nonumber\\
M^2_{\rm bos - bulk}&=& {n^2\over\rho^2}\label{mass}\\
M^2_{\rm boundary}&=& 0\ .\nonumber
\end{eqnarray}

Equation (\ref{mass}) shows that supersymmetry is broken, since
all supermultiplets in the bulk are split. In particular the 4D gravitino, as
well as all fermion moduli get a common mass:
\be
m_{3/2}={1\over 2\rho}\ .
\label{grav}
\en
This result provides the desired relation, discussed in section 3, so as to
describe gaugino condensation in the strong coupling regime of the heterotic
string. Moreover, since the observable sector lives on the boundary, 
we obtain from eq.~(\ref{mass}) vanishing
mass splittings to ${\cal O}(m_{3/2})$. Supersymmetry breaking will then be
communicated from the bulk by gravitational interactions, as was the case 
of the gaugino condensation models mentioned in the previous
section~\cite{gcold1,gcold2}.
 
The breaking (\ref{mass}) is identical 
to the one produced by finite temperature, 
where $\rho^{-1}$ is replaced by the temperature $T$, which plays the role 
of the inverse radius of the time coordinate. 
Since the tree-level scalar potential remains flat along the 
$\rho$-direction, the scale of supersymmetry breaking is
classically undetermined. To satisfy the unification condition (\ref{rho}) 
we need the gravitino mass to be at the intermediate scale $\sim 10^{12}$ GeV.
In the following, we will assume that this value is determined by some 
dynamical
mechanism in the M-theory analogue to the gaugino condensation, and explore
its consequences for supersymmetry breaking in the observable sector.

Note that the mass formula (\ref{mass}) is similar to the one obtained  in the
case of orbifold compactifications of the heterotic string,  where $\rho$ is
replaced by some large internal radius $R$~\cite{largeR}--\cite{largeR2}.  One
difference is the presence of the string winding modes whose inclusion is
determined by modular invariance.  Another substantial difference is the fact
that part of the observable matter, including all gauge multiplets, belong to
the untwisted (bulk) sector of the  orbifold. Thus, their infinite tower of
Kaluza--Klein excitations generically contributes to the renormalization of
gauge couplings  and lead to large threshold corrections, which break the
validity of  perturbation theory above the compactification scale $1/R$. One
then has to consider special models where this problem can be
avoided~\cite{modelsR}.  In contrast, this problem is absent in the present
case since only gravitationally coupled fields belong to the five-dimensional
bulk.

\nsect{Communication of supersymmetry breaking to the observable sector}

As explained in the previous section, at the lowest order, supersymmetry 
is broken only in the
five-dimensional bulk (gravitational and moduli sector), while it remains
unbroken in the observable sector. The communication of supersymmetry breaking
is then expected to arise radiatively, by gravitational interactions. This
issue is studied below.

\vspace{0.5cm}
\leftline{\large\it Scalar masses}

At the one-loop level, the diagrams that contribute to the scalar masses in the
observable sector are shown in Fig.~1, where the vertices come from the kinetic
terms. {}Fields from the boundary, denoted by $(\varphi,\psi_\varphi)$, 
always appear in pairs, 
as dictated by the $Z_2$ invariance. Moreover, in the effective 
field theory limit $\rho M_{11}\gg 1$, their couplings to fields from the bulk 
are the same for all Kaluza--Klein excitations. The latter are the moduli
$(z^{(n)},\chi^{(n)})$ and the graviton $(g_{\mu\nu}^{(n)},\psi_\mu^{(n)})$ 
supermultiplets. Moduli and matter field indices are dropped for notational
simplicity.

Every diagram from Fig.~1 gives a contribution proportional to:
\begin{eqnarray}
I=\kappa^2\sum_n\int{d^4k\over (2\pi)^4} {k^2\over k^2+M_n^2}
&=&\kappa^2\sum_n\int{d^4k\over (2\pi)^4} k^2
\int_0^\infty ds e^{-s(k^2+M_n^2)}\nonumber\\
&=&{1\over \pi M_p^2}\sum_n\int_0^\infty{ds\over s^3}e^{-s M_n^2}\ ,
\label{int}
\end{eqnarray}
where $\kappa=\sqrt{8\pi}/M_p$ and the masses $M_n$ are given in 
eq.~(\ref{mass}). After adding the contributions of diagrams related by
supersymmetry, one finds:
\begin{eqnarray}
I_{\rm bos}-I_{\rm ferm}&=&{1\over \pi M_p^2}\int_0^\infty{ds\over s^3}
\sum_n\left[ e^{-s n^2/\rho^2}-e^{-s (n+{1\over 2})^2/\rho^2}\right]\nonumber\\
&=&{1\over \pi M_p^2}\int_0^\infty{ds\over s^3}
\left[ \theta_3\left({is\over\pi\rho^2}\right)-
\theta_2\left({is\over\pi\rho^2}\right)\right]\nonumber\\
&=&{1\over \pi M_p^2}\int_0^\infty{ds\over s^3}
\left({\pi\rho^2\over s}\right)^{1/2}
\left[ \theta_3\left({i\pi\rho^2\over s}\right)-
\theta_4\left({i\pi\rho^2\over s}\right)\right]\ ,
\label{dint}
\end{eqnarray}
where $\theta_i$ are the Jacobi theta-functions and in the second line of
eq.~(\ref{dint}) we have used the Poisson resummation formula. Note that the
integral over $s$ is convergent both in the ultraviolet $(s=0)$ and in the
infrared $(s=\infty)$ regions.

After rescaling the integration variable $s=x\rho^2$, and
putting together all diagrams of Fig.~1, we obtain the following expression for
the scalar masses:
\be
m_{\varphi{\bar\varphi}}^2\sim G^{-1}_{\varphi{\bar\varphi}}\,
\left( G^{i{\bar\jmath}}R_{i{\bar\jmath}\varphi{\bar\varphi}}-
G_{\varphi{\bar\varphi}}\right)
{m_{3/2}^4\over M_p^2}{\cal J}\ ,
\label{mphi}
\en
where we used the relation (\ref{grav}) for $m_{3/2}$, and ${\cal J}$ is a 
constant given by:
\begin{eqnarray}
{\cal J}&=&\int_0^\infty{dx\over x^3}\left({\pi\over x}\right)^{1/2}
\left[ \theta_3\left({i\pi\over x}\right)-
\theta_4\left({i\pi\over x}\right)\right]\nonumber\\
&=&\sqrt\pi \int_0^\infty dy\; y^{3/2}\sum_n\left[1-(-)^n\right]
e^{-\pi^2n^2y}\label{J}\\
&=&{93\over 32\pi^4}\zeta(5)\simeq (0.17)^2\ ,\nonumber
\end{eqnarray}
where $y=1/x$ and $\zeta$ is the Riemann zeta-function. In eq.~(\ref{mphi}),
$G_{i{\bar\jmath}}$ and $G_{\varphi{\bar\varphi}}$ are the moduli and matter 
metrics, while 
$R_{i{\bar\jmath}\varphi{\bar\varphi}}$ is the moduli-matter mixed Riemann
tensor. The factor $G^{-1}_{\varphi{\bar\varphi}}$ comes from the wave
function renormalization and
the two terms in the bracket correspond to the contributions 
of the moduli and graviton supermultiplets, respectively, in the
loops of Fig.~1. 
As a result, we find the scalar masses 
$m_{\varphi{\bar\varphi}}={\cal O}(10^{-1})\; m_{3/2}^2/M_p\sim 10^4$~GeV 
generically. This is only a rough estimate, since besides the moduli 
dependent prefactor in eq.(\ref{mphi}), the result is very sensitive to the 
value of $M_{11}$. In fact, eqs.~(\ref{rel}) and (\ref{grav}) show that 
$m_{\varphi{\bar\varphi}}$ scales as
$M_{11}^6$ and, thus, a factor of 2 in $M_{11}$ changes the scalar masses by
almost two orders of magnitude.

Note that the scalar masses (\ref{mphi}) turn out to be diagonal in the flavour
space, if the matter metric is so. However, they need not be universal, owing
to the presence of the Riemann tensor which in general can introduce a
different moduli dependence for the various scalar fields. This may create
dangerous flavour-changing neutral currents, a problem that
is also shared by most string constructions and gives strong restrictions to
model building.

A similar analysis can be applied to compute the masses of the scalar moduli.
The corresponding diagrams are the same as those in Fig.~1, except that
$\varphi$ is replaced, 
in the external lines, by the lowest excitation of the moduli $z^{(0)}$ and
$(\varphi,\psi_\varphi)$ are replaced, in the internal lines, by the 
excitations $(z^{(-n)},\chi^{(-n)})$. 
Here, we use momentum conservation along the $x_5$-coordinate since
the vertices involve only fields from the bulk and can therefore be obtained 
by $Z_2$ truncation of the dimensionally reduced theory. The evaluation of the
diagrams can be done in the same way as before with the result:
\be
m_{z{\bar z}}^2\sim 5\, G^{-1}_{z{\bar z}}
\left( R_{z{\bar z}}-G_{z{\bar z}}\right)
{m_{3/2}^4\over M_p^2}{\cal J}\ ,
\label{mz}
\en
where $R_{z{\bar z}}$ is the moduli Ricci tensor and
the constant ${\cal J}$ is given in eq.~(\ref{J}). Thus, all moduli
scalars obtain masses of the same order as the scalar masses in the observable
sector ${\cal O}(10)$ TeV.

Although the moduli metric appearing in expressions (\ref{mphi}) and (\ref{mz})
is in general arbitrary, it has special properties in the limit under
consideration where the radius $\rho$ is large. In this limit, the theory 
behaves as five-dimensional with $N=2$ supersymmetry from the point of view of
four dimensions. The moduli metric then becomes block-diagonal in the space
of $N=2$ vector and hypermultiplets. {}Furthermore, in the large radius limit, 
the metric of the $h_{(1,1)}$ vector moduli is considerably simplified
by the underlying $N=2$ special geometry with a cubic
prepotential~\cite{ccaf}.
On the other hand, the metric of the remaining $h_{(2,1)}+1$ moduli, 
which can be obtained by a truncation of the corresponding quaternionic 
manifold, remains arbitrary in general. It would be interesting to further
analyse the consequences of these properties in particular models~\cite{lln}.

The fact that all scalar squared mass splittings are of order $m_{3/2}^4/M_p^2$
is a consequence of the absence of quadratic divergences in the effective
supergravity.\footnote{We thank S. Dimopoulos for discussions on this point.} 
Inspection of eq.~(\ref{dint}) shows that cancellation of 
quadratic divergences arises non-trivially. Indeed, any single
excitation $n$ of the sum gives a contribution to the integral, which is
quadratically divergent at $s=0$ as $ds/s^2$, so that after introducing an
ultraviolet cutoff $1/M_p^2$ one would get a contribution of
order $m_{3/2}$ to the mass. 
However, after summing over all modes and performing the
Poisson resummation, one finds that the integrand has an exponentially 
suppressed (non-analytic) ultraviolet behaviour as $e^{-\pi^2\rho^2/s}/s^{5/2}$.
This phenomenon is very similar to the situation at finite temperature upon the
identification $T\equiv 2m_{3/2}$, as was discussed in section 4. To make the
analogy more explicit, one can also compute the effective potential as a
function of the background $\rho$:
\begin{eqnarray}
V_{\rm eff}&=&{1\over 2}{\rm Str}\int{d^4k\over (2\pi)^4}
\ln (k^2+M^2)\nonumber\\
&=&-{N\over 32\pi^2}\int_0^\infty{ds\over s^3}\sum_n
\left[ e^{-s n^2/\rho^2}-e^{-s (n+{1\over 2})^2/\rho^2}\right]\nonumber\\
&=&-{N{\cal J}\over 32\pi^2}{1\over\rho^4}\ ,
\label{Veff}
\end{eqnarray}
where $N$ is the number of massless multiplets from the bulk. This result
corresponds to the well-known behaviour at finite temperature where 
$V_{\rm eff}\propto T^4$ and shows the vanishing of Str$M^2$ 
after supersymmetry breaking.

\vspace{0.5cm}
\leftline{\large\it Gaugino masses}

The diagrams that contribute to the gaugino masses at the one-loop level are
shown in Fig.~2, where the vertices again come from the kinetic terms. They
lead to individual contributions proportional to:
\be
\kappa^2\sum_n\int{d^4k\over (2\pi)^4} {|M_n|\over k^2+M_n^2}\ .
\label{intg}
\en
The total contribution to the mass can then be written as:
\begin{eqnarray}
m_{\lambda\lambda}&\propto&{1\over M_p^2}\int_0^\infty{ds\over s^{5/2}}
\sum_n\left[ e^{-s n^2/\rho^2}-e^{-s (n+{1\over 2})^2/\rho^2}\right]\nonumber\\
&\sim&{m_{3/2}^3\over M_p^2}\ ,
\label{mg}
\end{eqnarray}
where we followed the same steps as in the case of scalar masses.

The above result shows that the one-loop gravitational contributions to 
gaugino masses are too small for phenomenological purposes. This is a general
problem, which has been known for a long time~\cite{gcold1,gcold2}. 
A possible solution exists if there are massive
matter fields transforming non-trivially under the gauge group. Then, their
mass splittings generate gaugino masses by one-loop diagrams involving gauge
interactions. The latter lead to finite contributions given by~\cite{bgm}:
\be
m_{\lambda\lambda}\sim N_s
{\alpha\over 2\pi}\mu f\left({m_s\over\mu}\right)\ ,
\label{f}
\en
where $\mu$ is the supersymmetric mass and $m_s^2$ the squared mass 
splitting of those matter fields; $N_s$ denotes their multiplicity,
$\alpha$ is the corresponding gauge coupling, 
and the function $f(x)$ is nearly constant for $x\simgt 2$ 
while it behaves as $x$ for $x\simlt 1$. Thus, the gaugino masses are
of the order of $(\alpha/2\pi)N_s\, {\rm min}(\mu,m_s)$.

It is easy to see that when $\mu$ is below the intermediate scale $\rho^{-1}$, 
the evaluation of the scalar masses (\ref{int})--(\ref{mphi})
through the diagrams of Fig.~1 remains valid up to ${\cal O}(\mu/m_{3/2})$
corrections. It follows that the gaugino masses are approximately one order of
magnitude lower than the scalar masses if $\mu\simgt m_s$. Although 
this mechanism can give acceptable masses to charginos and neutralinos,
provided that the Higgs supersymmetric parameter $\mu$ is large enough, gluino
masses would require the Standard Model particle content to be extended by the
presence of extra colour multiplets in vector-like
representations such as triplets or leptoquarks. 
Of course, in this case, unification requires that the extra matter appears in
complete $SU(5)$ representations, e.g. $({\bf 5}+\overline{\bf 5})$ or 
$({\bf 10}+\overline{\bf 10})$.
Otherwise, in the absence of extra matter, this scenario
leaves open the possibility of having light gluinos with masses of order
$(\alpha_3/2\pi)\, m_{\rm top}={\cal O}(1)$ GeV.

To summarize, the mass spectrum we obtained in the observable sector 
originates from local supersymmetry breaking, with a gravitino mass $m_{3/2}$ 
at an intermediate scale defined by the size of the eleventh dimension of 
M-theory. All scalars then acquire masses of order 
$m_{3/2}^2/M_p$, while gaugino masses are of order $m_{3/2}^3/M_p^2$.
This situation is again identical to the case where supersymmetry is broken in
the heterotic string by gaugino condensation stabilized by a VEV of the 
antisymmetric tensor field strength~\cite{gcold1,gcold2}. 
As we saw, the problem of having very
light gauginos can be solved by means of gauge interactions involving extra
fields and providing gaugino masses of order $(\alpha/2\pi)m_{3/2}^2/M_p$.
Therefore, this scenario predicts a hierarchy of supersymmetric mass spectrum
where scalars are much heavier than gauginos. 

On the other hand, in the old analysis of gaugino condensation, 
based on the heterotic string tree-level effective supergravity, 
it was found that scalars remained massless at the one-loop order,
because of the dilation properties of the K{\"a}hler potential~\cite{gcold1}. 
In fact, it is easy
to see that the term in the brackets of eq.~(\ref{mphi}) vanishes when the
K{\"a}hler potential has for instance the no-scale $SU(1,n)$ form 
$K=-3\ln (z+{\bar z}- |\varphi|^2)$ and $\varphi$'s have zero VEVs.
This can lead to an alternative scenario where gauginos, with masses of order 
$m_{3/2}^3/M_p^2\sim 1$ TeV (for $m_{3/2}\sim 10^{14}$ GeV),
seed supersymmetry breaking in the rest of
the observable sector by gauge interactions. Since in this case the
corresponding diagrams are logarithmically divergent, all supersymmetric
masses turn out to be of the same order of magnitude. However, this scenario
is expected (and was explicitly shown~\cite{gcold2}) to be unstable 
under higher-order loop corrections, 
as the dilation symmetry is in general broken at the quantum level.

\nsect{Conclusions and outlook}

In this paper we have presented a mechanism for supersymmetry breaking by
compactification of M-theory to four dimensions. We argued that this mechanism
describes ordinary gaugino condensation in the strong coupling regime of the
heterotic string. It requires the eleventh dimension of M-theory to be large at
an intermediate scale $\rho^{-1}\sim 10^{12}$ GeV, which fixes the mass of the
gravitino as well as that of all fermionic partners of moduli that live on the
effective five-dimensional bulk. Supersymmetry breaking is communicated to our
observable world, which lives on the four-dimensional boundary, by 
gravitational
interactions. A computation of the mass splittings generically provides a
hierarchy of soft masses at the TeV scale $(\sim\rho^{-2}/M_p)$,
with matter scalars much heavier than gauginos.

Consistency of this mechanism requires the
vanishing of quadratic divergences, and hence the hierarchy of soft masses, 
to persist at higher loops. Indeed, by a preliminary inspection of 
higher-loop diagrams we expect this to be the case, in close analogy with the
situation at finite temperature. On the other hand, this scenario seems to
suffer from the usual cosmological problems, namely the cosmological constant
that appears to be of the order of the intermediate scale $m_{3/2}^4$ and the
problem of moduli that have masses at the TeV scale.
{}Furthermore, since scalar masses are not necessarily universal they might
induce flavour-changing neutral currents.
Although we have not discussed in this work the question of dynamical 
determination of the intermediate scale $\rho$, we would like to point out
that the required hierarchy $\rho M_{11}\sim 10^4$ is significantly smaller
than the usual hierarchy $M_G/m_{\rm SUSY}\sim 10^{13}$.

The above results can be contrasted with those where supersymmetry breaking is
induced by a large internal dimension $R$ 
in the heterotic string~\cite{modelsR,largeR2}. Unlike 
the present situation, in the latter case there is also, on top of the 
gravitational sector, ordinary matter including the gauge vector multiplets 
living in the bulk of the large internal dimension. As a consequence the
gravitino and gauginos get masses of the order of $R^{-1}$, and so the
corresponding radius $R$ is required to have a value of order 1 TeV$^{-1}$.
The seed of supersymmetry breaking is then communicated by gauge interactions
from the gauginos to the scalar sector. As a result, universality of scalar 
masses is approximately maintained and there is no hierarchy among 
soft-breaking parameters. The cosmological constant generated at 
one loop is ${\cal O}(R^{-4})$, much smaller than in the present mechanism. 
This also allows the dynamical determination of $R$ through the low-energy
radiative breaking of weak interactions. 
However, to avoid large threshold corrections to gauge couplings, spoiling the 
validity of perturbation theory, 
one has to consider special models where these corrections do not appear. 

We believe it is worth while to complete the present analysis 
with a more detailed 
computation of the mass spectrum and couplings (including trilineal soft terms)
of the low-energy effective field theory. In this work, we have used
a particular $Z_2$ discrete symmetry to break supersymmetry by the radius of
the eleventh dimension. One can thus ask the question of the existence of other
possible discrete symmetries that can also be used, and, of their effects in the
low-energy theory.\footnote{See also the recent analysis of Ref.~\cite{dg}.}
Nevertheless, the general features of the breaking will be
maintained, provided that one uses the same radius of $S^1$ to break 
supersymmetry in compactifications of M-theory on CY$\times S^1/Z_2$. The main
reason is that the observable sector lives on the boundary and, thus,
does not receive any mass splitting to lowest order. Supersymmetry breaking is
then communicated by gravitational interactions and yields the same hierarchy
in the mass spectrum as that we have found in this paper. The only possibility
to avoid this hierarchy would be in models with special dilation properties of
the K{\"a}hler potential. It would be interesting to study whether such models
can be built using the particular features of the effective theory in the
large radius limit discussed in section 5, and whether these models can 
escape some of the above-mentioned problems.
{}Finally, it is an open question to find
an M-theory formulation of this mechanism of supersymmetry breaking
using for instance M(atrix)-theory.

\newpage
\begin{figure}
\centerline{
\psfig{figure=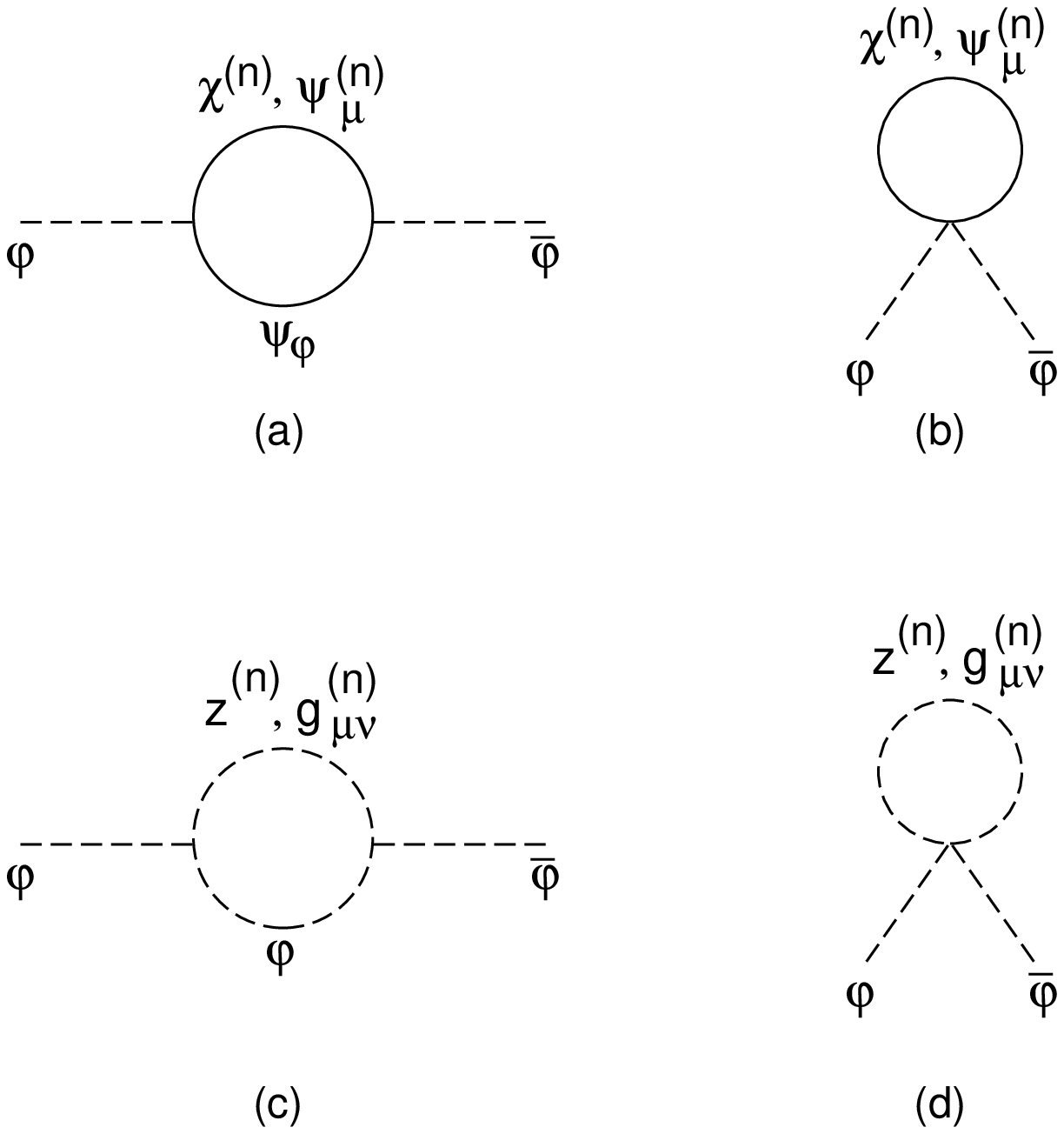,width=12cm,height=12.0cm}}
\caption{One-loop (gravitational) diagrams contributing to scalar masses.}
\label{feynman1}
\end{figure}
\begin{figure}
\centerline{
\psfig{figure=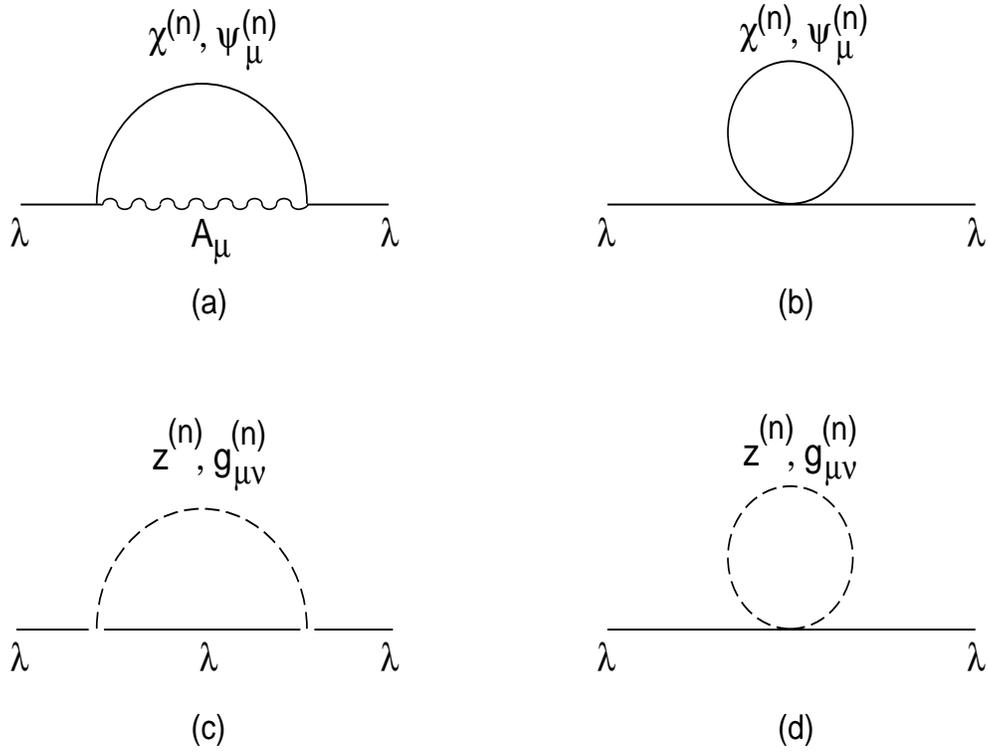,width=13cm,height=10.0cm}}
\caption{One-loop (gravitational) diagrams contributing to gaugino masses.}
\label{feynman2}
\end{figure}

\end{document}